\documentclass[11pt,twoside]{article}
\usepackage{asp2010}
\usepackage{graphicx}
\usepackage{epstopdf}

\resetcounters

\begin{document}

\title{Helioseismic detection of emerging magnetic flux}
\author{Stathis I\scriptsize{LONIDIS}\normalsize, Junwei Z\scriptsize{HAO}\normalsize, and Alexander G. K\scriptsize{OSOVICHEV}\normalsize}
\affil{W.W. Hansen Experimental Physics Laboratory, Stanford University, Stanford, CA 94305-4085, USA}

\begin{abstract}
Investigating the properties of magnetic flux emergence is one of the most important problems of solar physics.
In this study we present a newly developed deep-focus time-distance
measurement scheme which is able to detect strong emerging flux events in the
deep solar interior, before the flux becomes visible on the surface. We discuss in detail the differences
between our method and previous methods, and demonstrate step-by-step how the
signal-to-noise (S/N) ratio is increased. The method is based on detection of perturbations in acoustic phase travel times determined from cross-covariances of solar oscillations observed on the surface. We detect strong acoustic travel-time reductions of an order of $12-16$ seconds at a depth of $42-75$ Mm. These acoustic anomalies are detected $1-2$ days before high peaks in the photospheric 
magnetic flux rate implying that the average emerging speed is $0.3 - 0.6$ km s$^{-1}$. The results of this work
contribute to our understanding of solar magnetism and benefit space weather forecasting.
\end{abstract}

\section{Introduction}
\par Solar magnetic fields are presumably generated by a dynamo action
in the deep interior of the Sun and emerge through the convection zone to the
photosphere. The properties of magnetic flux emergence are related to some 
of the most important problems of solar physics: the depth of dynamo, 
the appearance and evolution of active regions, the formation of sunspots, 
the initiation of flares and coronal mass ejections, and the
11-year activity cycle. The detection of emerging magnetic flux events in the
deep solar interior may improve our understanding of solar magnetism.
 
\par Early detection of emerging magnetic structures in the solar interior will also benefit
space weather forecasting. Sunspot regions produce flares and coronal mass ejections (CMEs)
which may cause power outages as well as interruptions of telecommunication and navigation
services. It is important therefore
to monitor the subsurface magnetic activity and, if it is possible, to predict the emergence
of sunspots as well as the eruptive events associated with them. 
 
\par There have been several attempts to detect emerging magnetic flux prior to its
appearance in the photosphere. 
Chang et al. (1999) constructed phase-shift maps of
Active Region (AR) 7978 using the method of acoustic imaging (Chang et al. 1997) and reported the
detection of upward-moving magnetic flux during the development of
the active region. Kosovichev et al. (2000) studied, with the time-distance technique (Duvall et al. 1993), the emergence of an active
region which appeared on the solar disc on 1998 January 11 and estimated the emerging speed at about $1.3$ km s$^{-1}$. Jensen
et al. (2001) analyzed the same active region and found that
wave-speed perturbations extend at least $20$ Mm below the active
region.
Kosovichev \& Duvall (2008) investigated the emergence of AR 10488
which was observed with the Solar and Heliospheric Observatory (SOHO)
Michelson Doppler Imager (MDI) in October 2003. The authors
suggest that the active region was formed as a result of multiple
flux emergence events. They also found that the emergence was
accompanied by strong shearing outflows. Zharkov \& Thompson (2008)
investigated the emergence of AR 10790 using the same time-distance approach as Kosovichev et al. (2000). They observed regions
with sound speed perturbations, presumambly related to subsurface
emerging flux, and estimated the speed of emerging flux at
about $1$ km s$^{-1}$. Komm et al. (2009) found, using a
ring-diagram analysis (Hill 1988), that at some depth ranges, the vertical flows
show temporal variations that correspond to photospheric variations of the magnetic flux. 
Hartlep et al. (2011) showed that, under certain conditions, subsurface structures
related to emerging magnetic flux can modify the acoustic power
observed at the photosphere above them.
For a recent review on the dynamics of emerging magnetic flux 
see Kosovichev (2009).

\par Recently, Ilonidis et al. (2011) detected significant travel-time 
perturbations at a depth of about $42-75$ Mm and 
showed that these perturbations were associated with magnetic structures that
emerged with an average speed of $0.3-0.6$ km s$^{-1}$ and appeared at the surface 
$1-2$ days after the detection of the perturbations. Here, we describe the analysis method in detail and
demonstrate step-by-step how the S/N ratio is increased. We also present travel-time maps
of 2 emerging flux regions which show the temporal evolution of the detected travel-time anomalies.

\section{Data and Method}
\par Doppler observations from MDI onboard SOHO (Scherrer et al. 1995) are used in this
work. Each dataset is 8-hours long, tracked with a Carrington
rotation rate and remapped to heliographic coordinates using Postel's projection. The datasets are
tracked with a spatial resolution of $0.12^\circ$ pixel$^{-1}$, a temporal
resolution of 1 minute, and a
size of $256 \times 256$ pixels. Each dataset is then filtered in the Fourier
domain and only the oscillation signals with frequencies between
$2-5$ mHz and phase speeds between $92-127$ km s$^{-1}$ are kept for
further processing.

\begin{figure}[h!]
\centering
\includegraphics[width=1.\textwidth]{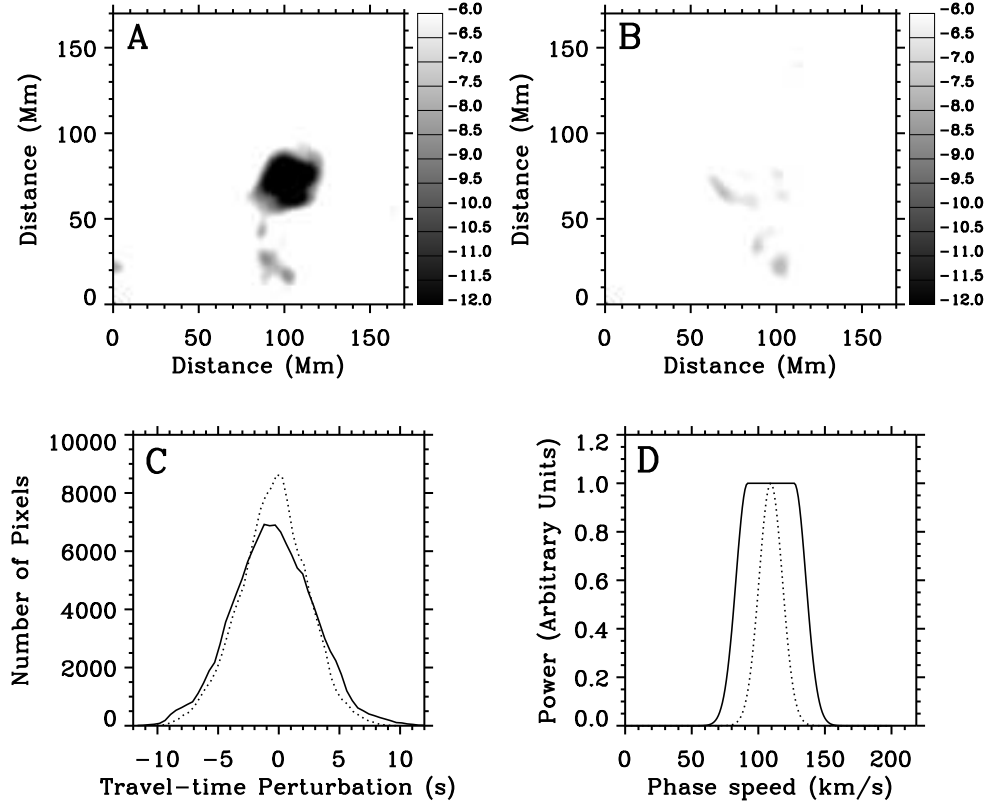}
\vspace{-2.5pc}
\caption{\label{label1}(A)  Mean travel-time perturbation map (in seconds) 
of AR 10488 at a depth of $42-75$ Mm obtained from an 8-hour dataset centered at 03:30 UT, 26 October
2003. The map was computed using the phase-speed filter shown in panel D with solid line and arcs with 
a size of $45^{\circ}$ and 4 different orientations. The maximum phase travel-time shift is 16.2 s.
(B) Same as panel A except that the map was computed
using the Gaussian phase-speed filter shown in panel D with a dotted line. The maximum travel-time shift is 7.4 s. (C)  Distribution of the travel-time shifts measured in 9 quiet-Sun regions using exactly the same procedure as in panel A (solid line) and exactly the same procedure as in panel B (dotted line). The standard deviations of these measurements is 3.4 s and 2.9 s which yield S/N ratios of 4.8  and 2.6 for the
signature detected in panels A and B respectively. (D) The two phase-speed filters used for computation of the travel-time maps in panels A and B.}
\end{figure}

\begin{figure}[h!]
\centering
\includegraphics[width=1.\textwidth]{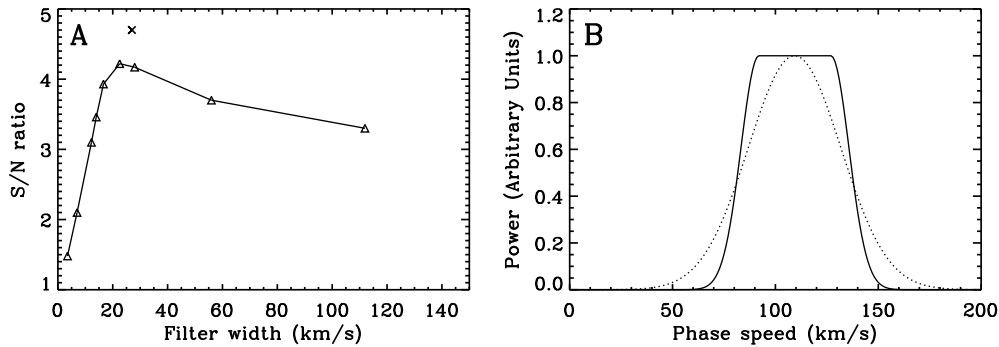}
\caption{\label{label1} Panel A shows the S/N ratio as a function of the width (standard deviation) of a Gaussian phase-speed filter.  The S/N ratios were estimated by analyzing emerging-flux and quiet-Sun regions, as shown in Fig. 1, using the method described there. The cross sign ($\times$) indicates the S/N ratio of the optimized filter shown in panel  B with a solid line which was used for the detection of emerging sunspot regions. This filter has higher S/N ratio than any Gaussian phase-speed filter with the same central phase-speed. The dotted line in panel B shows the Gaussian filter which yields the highest S/N ratio among the Gaussian filters of panel A.}
\end{figure}

\begin{figure}[h!]
\centering
\includegraphics[width=1.\textwidth]{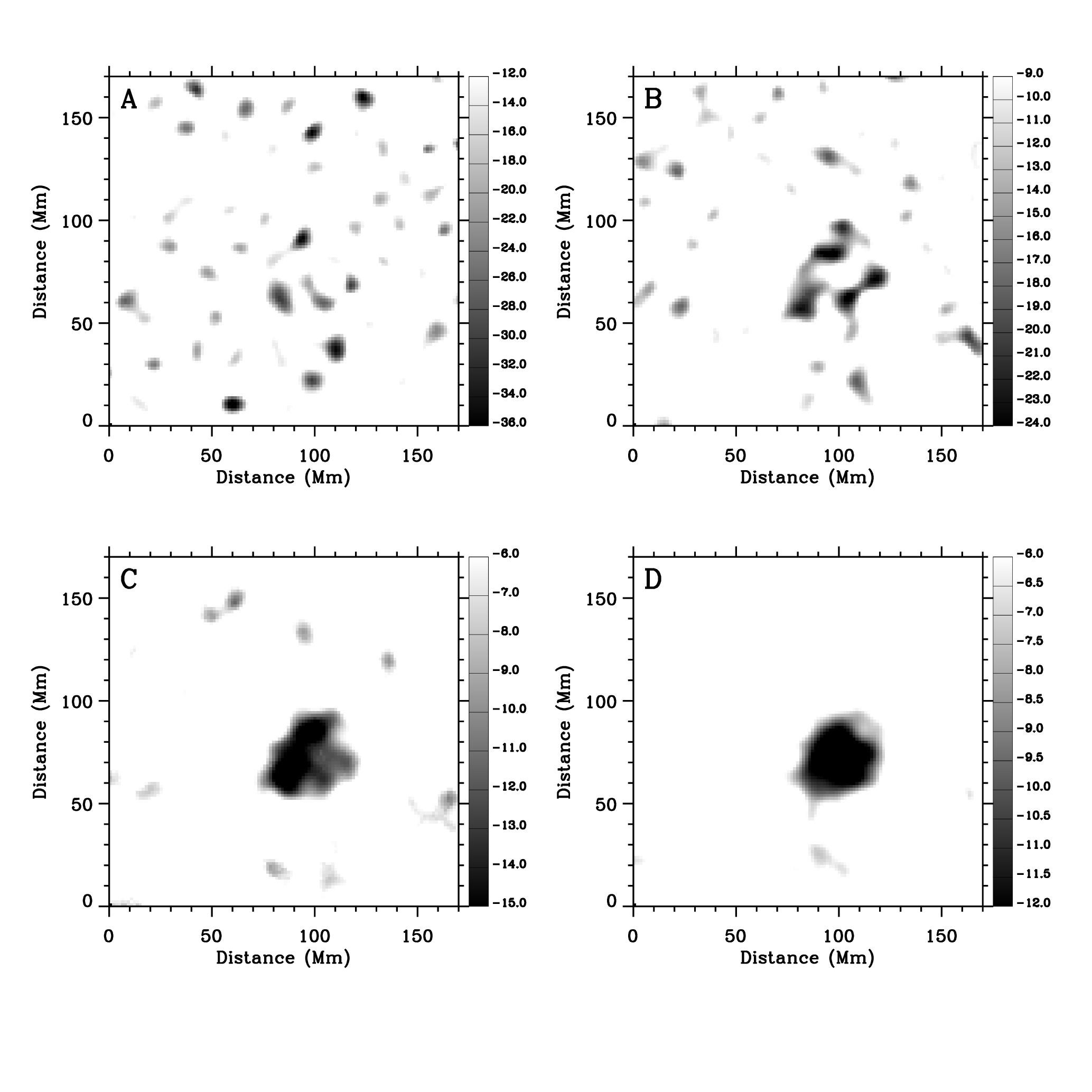}
\vspace{-2.5pc}
\caption{\label{label1} Demonstration of a step-by-step improvement of the S/N ratio in the mean travel-time perturbation maps. The emerging-flux region is the same as in  Fig. 1. Each panel introduces one or more modifications to the previous deep-focus time-distance measurement scheme: (A) The previous method: point-to-point computation of cross-covariance function with a Gaussian phase-speed filter with width (standard deviation) of 12.3 km s$^{-1}$ using 9 focus depths between 54 and 62 Mm. The S/N ratio is rather low. (B) Same as in panel A but with 2 modifications: the map was computed using the new filter (solid line of Fig. 2 B) and 31 focus depths between 42 and 75 Mm. The S/N ratio is 3.2. (C) Same as in panel B but with one additional modification: the cross-covariances are computed after averaging the oscillation signal over arcs with a size of $25.7^{\circ}$. The S/N ratio is 3.9. (D) Same as in panel C but with one additional modification: the cross-covariances are computed using the 20 arc configurations described in the text. The S/N ratio is 4.9.}
\end{figure}

\par A new deep-focus time-distance measurement scheme is then applied
for the computation of the cross-covariances. This measurement scheme introduces
4 modifications to the deep-focus time-distance measurement schemes that
were used in previous studies. We discuss
these modifications below and we show that they improve
the S/N ratio.

\par The first modification is related to the phase-speed filtering. Phase-speed filters select acoustic 
waves with the same (or similar) phase speed(s). These waves have the same penetration depth and can be used
to investigate properties of the flux emergence at specific depths. The most challenging part in
the selection of a phase-speed filter is the determination of the exact shape and width of the filter.
Helioseismology usually uses Gaussian phase-speed filters with the width chosen
empirically. However, the choice of the shape and/or the width of the phase-speed filter can affect both 
the signal (travel-time shift 
caused by an emerging flux event) and the noise (travel-time shifts measured in quiet-Sun regions) levels.
Figure 1 shows an example where two phase-speed filters with the same central phase speed are tested
in the same emerging-flux and quiet-Sun regions. It turns out that the filter with the lower noise level, estimated
by the quiet-Sun measurements,
does not have a higher S/N ratio because of the significant reduction on the signal level. This simple example
demonstrates the importance of the optimal phase-speed filter selection. Gaussian phase-speed filters that minimize the noise
level and have been widely used in helioseismology may not necessarily maximize the S/N ratio.
Here, we choose a $\Pi$-shaped filter which selects all the acoustic waves with phase
speeds between $92-127$ km s$^{-1}$ and drops, outside of this range, as a Gaussian function with width (standard deviation) 
of 8.7 km s$^{-1}$. The S/N ratio of this phase-speed filter is higher than any Gaussian phase-speed filter with the same central phase-speed (Fig. 2).  

\par The second modification is related to the computation of cross-covariances. 
According to the previous deep-focus time-distance measurement scheme, an annulus
is selected on the solar surface and the oscillation signals at every pair of diametrically opposite 
points on this annulus are cross-correlated both for positive and negative time lags. The
cross-correlation functions are then averaged and the two lags are combined to
increase the S/N ratio. Here, we follow a different approach. We select an annulus on the
solar surface and devide it into an even number of arcs. We average the oscillation signal
inside each arc and compute cross-covariances between the averaged signals of every pair 
of diametrically opposite arcs. The cross-covariance functions are computed for both positive and negative
time lags. These functions are then averaged and the two lags are combined to increase the S/N ratio.
Ilonidis and Zhao (2011) used this procedure, with an  arc size of  $90^\circ$, to measure the coefficients of absorption, 
emissivity reduction, and local suppresion inside sunspots.

\par Averaging the oscillation signal over an arc allows us to use many different arc
configurations for the computation of cross-covariances. This is the third modification to the previous 
measurement scheme. In the new scheme, both the arc size and the arc orientation
are free parameters that can be chosen to maximize the S/N ratio. We select arcs of 5 different 
sizes and for each arc size, we use 4 orientations. The arcs have a size of $25.7^\circ$, $30^\circ$, 
$36^\circ$, $45^\circ$ and $60^\circ$ (so that the annuli are divided into 14, 12, 10, 8 and 6 arcs respectively).
We select the 4 orientations starting from an arbitrarily oriented configuration with N arcs in such a way that every other
configuration with N arcs is produced from the previous one by a rotation of N/4 degrees. 
In total, we use 20 different arc configurations. The cross-covariances computed from all these configurations
are combined to increase the S/N ratio. 

\par The last modification is related to the depth range of our measurements. The goal of a 
helioseismic study is often to map perturbations in the solar interior. The oscillation signals which
are selected and cross-correlated on the solar surface correspond to the starting and ending points
of acoustic wave paths which, in the ray-path approximation, are focused on a single point deep below the photosphere. Therefore,
perturbations in the acoustic travel time essentially map local inhomogeneties around the focal point. Here,
we combine the
cross-covariances obtained from 31 travel distances (corresponding to 31 focus depths) in order to
increase the S/N ratio. The total depth range of our measurements is $42-75$ Mm. So the travel-time maps
presented here are sensitive to
acoustic anomalies in this wide range of depths. We should note that cross-covariances obtained for
the different travel distances are averaged after 
appropriate time shifts based on quiet-Sun measurements.

\par We compute the cross-covariances using the method described above with the new optimal phase-speed filter, the 20 arc
configurations and the 31 travel distances, and we combine these cross-covariances in order to
increase the S/N ratio. The final cross-covariance is fitted with a Gabor wavelet
(Kosovichev and Duvall 1996) to obtain the acoustic phase travel time of one cross-covariance peak. This procedure is repeated for all the pixels in the observed area, and an
acoustic phase travel-time map is constructed. We should note that the definition of phase travel time may be different in other local helioseismology methods and this difference should be taken into account when comparing results obtained with different methods. 

\section{Results}

\par We present travel-time maps of two emerging flux regions. Active Region (AR) 10488 was one of the largest active regions of Solar Cycle 23. It started emerging on the solar disc at 09:30 UT, 26 October 2003, about $30^{\circ}$ East of the central meridian. The emergence showed a steep increase in the magnetic flux rate with a high peak at about 08:00 UT, 27 October. Travel-time maps of this active region at a depth of about $42-75$ Mm are shown in Fig. 4. A strong subsurface travel-time anomaly started developing at about 23:30 UT, 25 October (the time always correponds to the mid-point of an 8-hour dataset used for cross-covariance computations), 10 hours before the start of emergence. This feature increased in size and strength during the next $4-5$ hours and then gradually weakened, over a period of 3-4 hours, until the signal fell below the noise level. This feature had a maximum travel-time perturbation, estimated by our method, of 16.3 s relative to the quiet Sun. No other strong travel-time anomalies were detected at this location during the next $1-2$ days. It should be pointed out that the development and decay of this strong subsurface travel-time anomaly happened several hours before the start of emergence at the surface. The magnetic field at that time and until the start of emergence was very quiet, and no signature of magnetic field emergence was visible in the magnetograms. Magnetic field observations as well as plots of the total flux and the flux rate of this active region can be found in Ilonidis et al. (2011).

\begin{figure}[h!]
\centering
\includegraphics[width=1.\textwidth]{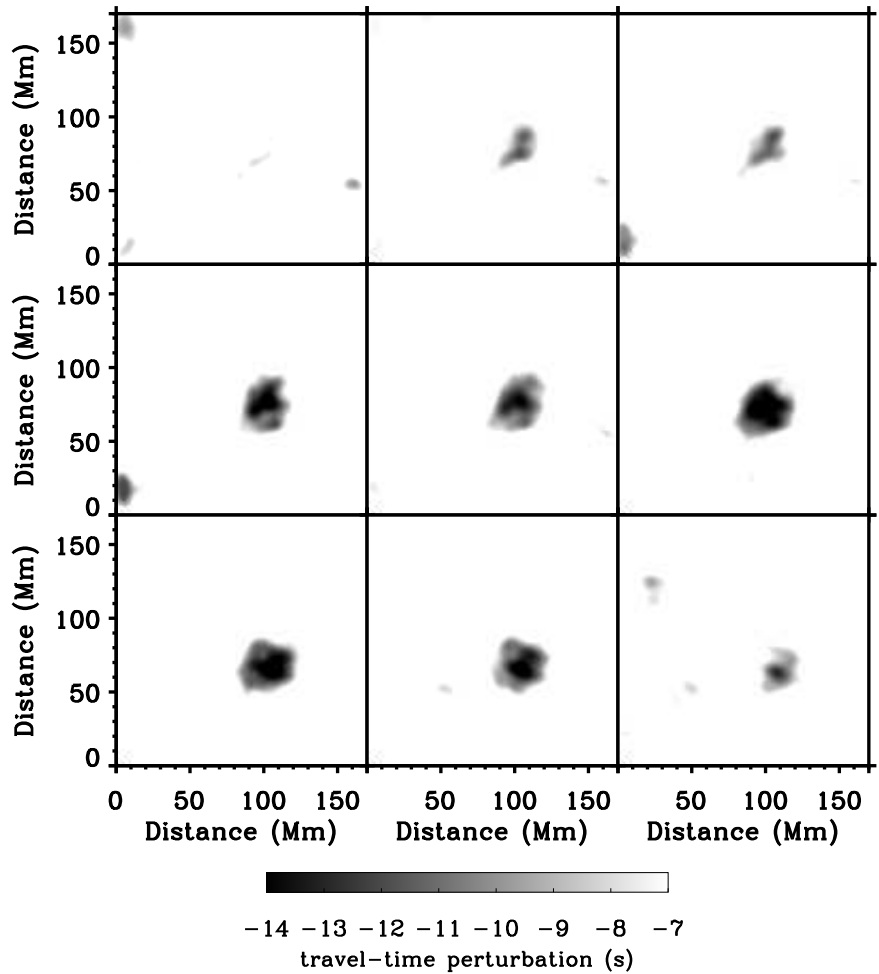}
\caption{\label{label1}  Mean travel-time perturbation maps 
of AR10488 at a depth of $42-75$ Mm obtained from 8-hour MDI Doppler-shift datasets. From top left to bottom right, the maps are centered at 22:30 UT 25 October 2003, 23:30 UT, 00:30 UT 26 October 2003, 01:30 UT, 02:30 UT, 03:30 UT, 04:30 UT, 05:30 UT, 06:30 UT.  A strong travel-time anomaly appeared before the start of emergence and persisted for about 8 hours.}
\end{figure}

\begin{figure}[h!]
\centering
\includegraphics[width=1.\textwidth]{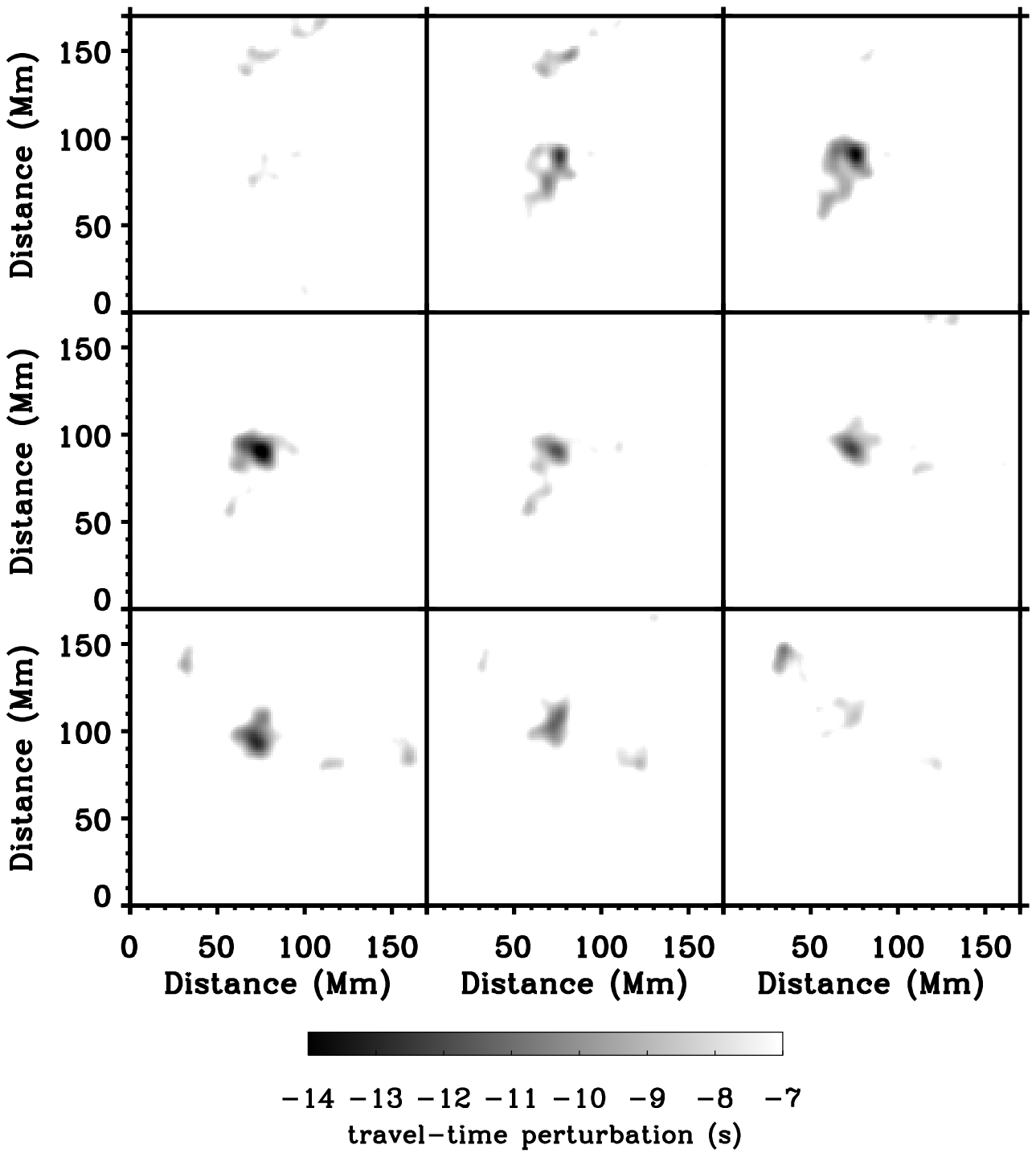}
\caption{\label{label1}  Mean travel-time perturbation maps 
of AR 8164 at a depth of $42-75$ Mm obtained from 8-hour MDI Doppler-shift datasets. From top left to bottom right, the maps are centered at 22:00 UT 22 February 1998, 23:00 UT, 00:00 UT 23 February 1998, 01:00 UT, 02:00 UT, 03:00 UT, 04:00 UT, 05:00 UT, 06:00 UT. A strong travel-time anomaly appeared before the start of emergence and persisted for about 7 hours.}
\end{figure}

\par AR 8164, which was smaller and less active than AR 10488, started emerging in the northern solar hemisphere at about 04:00 UT, 23 February 1998. The magnetic flux rate at the surface showed initially a steep increase followed by a high peak at about 08:00 UT, 24 February. Figure 5 shows travel-time maps of this active region before and during the start of emergence. A strong travel-time anomaly appeared first in the travel-time maps at 23:00 UT, 22 February and persisted for about 7 hours. This feature had a maximum travel-time perturbation of about 14.0 s. 

\par Our analysis method was also applied to 9 quiet-Sun regions with no emerging flux events in order to estimate the noise level of our measurements. The analysis method (measurement scheme, phase-speed filtering, fitting etc) was exactly the same as for the emerging flux regions. The quiet-Sun measurements showed no significant travel-time perturbations. The standard deviation of these measurements was 3.3 s which yields a S/N ratio for ARs 10488 and 8164 of 4.9 and 4.2 respectively. The travel-time shifts measured in quiet regions can be caused by the stochastic realization noise of solar oscillations, small errors in the fitting of the cross-covariance function, or by physical effects such as thermal perturbations and weaker magnetic fields in the convection zone.

\par The detection of the strong signals in Figs. 4 and 5, which persist for about 8 hours, does not necessarily mean that the flux emergence takes place for only 8 hours. Perhaps a more appropriate interpretation is that the flux is strong enough to cause detectable travel-time shifts by this method only for about 8 hours. The detected travel-time perturbations correspond, most probably, to strong emerging flux events at a depth of $42-75$ Mm which reach the surface $1-2$ days after the detection and cause high peaks in the photospheric flux rate. The magnetic flux observations at the surface and our helioseismic measurements confirm this scenario. The strongest travel-time anomalies of ARs 10488 and 8164 were detected about 28.5 and 32 hours respectively before the highest peaks in the corresponding flux rates. The magnetic field of AR 10488, which was stronger (and therefore more buoyant) than AR 8164 caused larger travel-time shifts and emerged faster in the photosphere. The study of 2 other emerging flux events, presented in Ilonidis et al. (2011), confirm this scenario as well. The estimated average emerging speed of about $0.3 - 0.6$ km s$^{-1}$ is consistent with estimates from numerical simulations (Fan, 2009).

\par The horizontal wavelength of the acoustic waves employed in this study is about 35 Mm at 3.5 mHz. The large wavelength poses limits on both the size of structures that can be resolved and the accuracy of their location. The amplitude of the travel-time perturbations is larger than theoretical estimates based on numerical simulations of emerging flux tubes (Birch et al. 2010). These estimates were derived though with a very different method, and at this point it is not known if the discrepancies are caused by the differences in the two methods. In addition, the interpretation of the detected signals is not a simple task because the nature of the travel-time anomaly is not known. Observational studies on the interaction of acoustic waves with magnetic fields (Zhao 2011) show that magnetic fields cause complicated changes in the cross-covariance function and not simple uniform shifts. Perhaps, a more detailed study of the cross-covariance function may give some hints about the nature of the detected perturbations.

\par Monitoring and predicting solar magnetic activity is a useful tool for space weather forecasts. Sunspot regions can either emerge from the solar interior or rotate into our view from the Sun's East limb. Our method combined with continuous observations from the Helioseismic and Magnetic Imager onboard the Solar Dynamics Observatory and the far-side imaging technique (Lindsey \& Braun 2000; Zhao 2007; Ilonidis et al. 2009) may allow anticipation of large sunspot regions days before they appear on the solar disc. 

\acknowledgements The authors thank P. Scherrer and T. Duvall for discussions and useful comments.


\end{document}